\documentstyle[aps,psfig]{revtex}
\begin{document}

\begin{title}
{
\hfill{\small {\bf MKPH-T-01-09}}\\
{\bf The two-nucleon system in the $\Delta$ region
      including full meson retardation} \footnote[2]
{Supported by the Deutsche Forschungsgemeinschaft (SFB 443).}}
\end{title}

\author
{Michael Schwamb and Hartmuth Arenh\"ovel}
\address{Institut f\"ur Kernphysik,           
         Johannes Gutenberg-Universit\"at,  
         D-55099 Mainz, Germany }
\maketitle

\begin{abstract}
A model is developed for the hadronic and electromagnetic interaction
 in the two-nucleon system above pion threshold in the framework of meson,
 nucleon and $\Delta$ degrees of freedom. It is based on time-ordered
 perturbation theory and  includes full meson
 retardation in potentials and exchange currents
  as well as loop contributions to the nucleonic one-body current. 
 Results for $NN$ scattering and  deuteron photodisintegration 
 are presented.
\end{abstract}

\section{Introduction}
 At present, a very interesting topic in the field of medium energy physics
is devoted to the role of effective degrees of freedom (d.o.f.)
 in hadronic systems in  terms of nucleon, meson and isobar d.o.f.
 and their connection to the underlying quark-gluon dynamics of
 QCD.  For the study of 
this basic question, the two-nucleon system provides an 
important test laboratory, because it is  the simplest nuclear 
system for the study of the nucleon-nucleon interaction.  Moreover,
  the role of medium effects due to two-body 
 operators and the role of offshell effects, i.e.\ the change of single
 particle properties in the nuclear medium, can be investigated most
 precisely.  

State-of-the-art  models for describing  hadronic and electromagnetic
 reactions on the two-nucleon system up to the $\Delta$ region
 should incorporate
 -- among other things -- a dynamical treatment of the $\Delta$ isobar.
 Moreover, gauge invariance and unitarity should be fulfilled at least
 approximately.  Within a unitary model,  
 the various possible  reactions  cannot be treated 
 independently, because they are linked by the optical theorem.
 For example, for energies up to the two-pion threshold 
 the forward Compton scattering amplitude is related via
\begin{eqnarray}\label{opt_th}
 \mbox{Im} \, T(\gamma  d \rightarrow \gamma  d;\theta =0) & \, \sim \, & 
   \sigma_{tot} (\gamma  d \rightarrow  NN, \pi d, \pi NN)
\end{eqnarray}
to the sum of the cross sections of photodisintegration and coherent and
incoherent pionphotoproduction. Therefore, all reactions on 
 the two-nucleon system  should be described within one consistent
 framework. In the past years, we have started to realize
 this ambitious project  within a {\it retarded}
 coupled channel $NN/N\Delta$-approach based on three-body scattering
 theory with nucleon, $\Delta$ and meson degrees of freedom
\cite{ScA98a,ScA98b,ScA01a,ScA01b,ScA01c}.

\section{The model}

 In order  to motivate why retardation should be taken
 into account above     pion threshold, let us consider
 an arbitrary two-body meson-exchange operator like the ordinary
 one-pion exchange potential.
In time-ordered perturbation theory, the propagation of the intermediate
 $\pi NN$ system  is described by
 the retarded propagator 
 $G_0^{ret}(E +i \epsilon)=\left(E+i\epsilon -H_0\right)^{-1}$,
where $E$ is the invariant energy of the system and $H_0$ denotes the 
 kinetic Hamilton operator for the intermediate $\pi NN$ system.
  Due to its nonhermiticity, nonlocality
 and the existence of singularities above pion threshold,
 an exact treatment of $G_0^{ret}$ is quite complicated. Therefore, 
 in most practical
 applications a  low energy approximation, the so-called
 {\it static limit} is used by neglecting the energy transfer between
 the nucleons by the pion. The corresponding
 propagator $G_0^{stat}$
 is much easier to handle. One encounters on the other hand
 at least two serious problems. Above
 pion-threshold, unitarity is violated due to the absence of  
 singularities in $G_0^{stat}$.  Moreover, in the past it turned out 
 that even the simplest photonuclear reaction, namely
 deuteron photodisintegration, cannot be  described even
 qualitatively within a consistent
 static framework \cite{Ta089,WiA93}.

For incorporating retardation, the mesons generating the nucleon-nucleon
 interacton and the meson-exchange currents  have to be treated
 as explicit degrees of freedom. 
Therefore, 
the model Hilbert space consists then of three orthogonal subspaces
${\cal H}^{[2]} = {\cal H}^{[2]}_{\bar N} \oplus
{\cal H}^{[2]}_{\Delta} \oplus 
{\cal H}^{[2]}_{X}\, $, 
where ${\cal H}^{[2]}_{\bar N}$ contains two bare nucleons, 
${\cal H}^{[2]}_{\Delta}$ one nucleon and one $\Delta$ resonance, 
and ${\cal H}^{[2]}_{X}$ two nucleons and one meson 
$X \in \{ \pi, \rho, \sigma, \delta, \omega, \eta\}$.
Concerning the hadronic part,  the basic  interactions in our model 
 are $XNN$  and $\pi N \Delta$ vertices.
   Inserting these into the Lippmann-Schwinger equation,
 one obtains after some straightforward algebra \cite{ScA01a} effective
 hadronic  interactions acting in 
 ${\cal H}^{[2]}_{\bar N} \oplus  {\cal H}^{[2]}_{\Delta}$ which contain
 the desired
 {\it retarded} one-boson exchange (OBE) mechanisms describing the transitions
 $NN \rightarrow NN$, $NN \rightarrow N\Delta$ and  
 $N\Delta \leftrightarrow N\Delta$. This strategy of starting
 with vertices as basic interaction terms  in the enlarged
 Hilbert space ${\cal H}^{[2]}$
 has the advantage that no inconsistencies occur. In this context, note
 especially that the vertices 
 are hermitean. On the other hand, if one started with a retarded OBE
 as basic interaction
 in  ${\cal H}^{[2]}_{\bar N} \oplus  {\cal H}^{[2]}_{\Delta}$,
  one would loose hermiticity and therefore the
 solid grounds of quantum mechanics.

In  our explicit realization, we use for the parametrization of the
 retarded $NN$ interaction
  the Elster potential \cite{ElF87} which takes into account
 in  addition one-pion loop
 diagrams  in order to fulfill unitarity above pion threshold. Therefore,
 one has to distinguish between bare and physical nucleons
 (see \cite{ScA01a} for details). Concerning the
 transitions $NN \rightarrow N\Delta$ and 
 $N\Delta \leftrightarrow N\Delta$, we take besides retarded
 pion exchange   static $\rho$ exchange into account. Moreover,
 the interaction of two nucleons in the deuteron channel in presence of
 a spectator pion (the so-called $\pi d$ channel) is also considered. By a
 suitable box renormalization \cite{GrS82}, we are able to obtain approximate
 phase equivalence between the Elster potential and our coupled
 channel approach below pion threshold.
 
Similarly, the basic electromagnetic interactions  consist of
 baryonic and mesonic one-body currents as well as  vertex and 
 Kroll-Rudermann contributions \cite{ScA01b,ScA01c}. These currents are,
 together with the $\pi NN$ vertex, the basic building blocks
 of the corresponding effective
 current operators. The latter contain beside the ordinary spin-,
 convection- and spin-orbit current full retarded pionic 
 meson exchange currents
 and electromagnetic loop contributions, where the latter
 can be interpreted as  off-shell contributions  to the baryonic
 one-body current \cite{ScA01c}. Moreover, static $\rho$ MEC as well as
 $\Delta$ MEC contributions are taken into
 account. It can be shown \cite{ScA01c} that concerning the pionic part
 gauge invariance is fulfilled  in leading order of $1/M_N$.

\section{Results}
\begin{figure}[hp]
\centerline{\psfig{figure=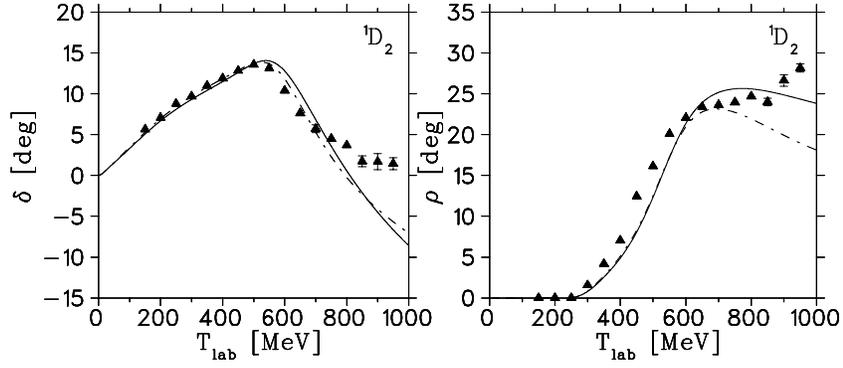,width=11cm,angle=0}}
\vspace{0.5cm}
\caption{Phase shift $\delta$ and inelasticity $\rho$ for the
 $^1D_2$ $NN$-channel in comparison with experiment (solution SM97 of
 Arndt {\it et al.}~{\protect\cite{Arn98}})
for two  potential models: dash-dotted curve: static approach, based
 on the Bonn-OBEPR potential {\protect\cite{MaH87}},
 full curve: retarded approach. See {\protect\cite{ScA01a}} for further
 details.}
\label{nnphase}
\end{figure}

The hadronic $\Delta$ parameters and the M1 $\gamma N \Delta$ coupling
 are simultaneously fitted  to the $M_{1+}(3/2)$
 multipole of pion  photoproduction on the nucleon,  the $P_{33}$ channel
 in pion-nucleon scattering and the $^1D_2$ channel in  nucleon-nucleon
 scattering \cite{ScA01a,ScA01b}. In Fig.\ \ref{nnphase}, our results for the
 $^1D_2$ phase shift and inelasticity are depicted. 
 We obtain a good description at least up to  about $T_{lab}=800$ MeV. 
 Concering the other partial $NN$ waves, the overall description
 is fairly well but needs some further improvement in the future 
 \cite{ScA01a}. Therefore, at present
 we are  constructing from scratch an improved hadronic interaction 
 model whose parameters are fitted to the phase shifts and inelasticities
 of {\it all} relevant $NN$ scattering partial waves for $T_{lab}$ energies
 up to about 1 GeV.

\begin{figure}[hp]
\centerline{\psfig{figure=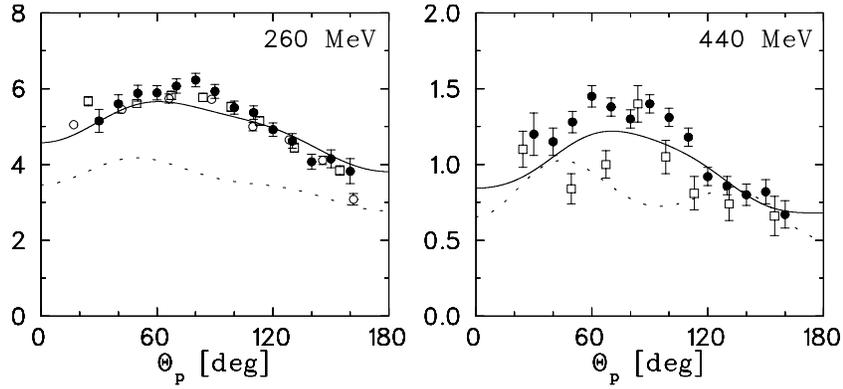,width=11cm,angle=0}}
\vspace{0.5cm}
\caption{Differential cross section of deuteron photodisintegration
 for two photon energies $k_{lab}$ as  function
 of the c.m.\ proton angle $\theta_p$: dotted curve: result of
 Wilhelm {\it et al.} in static approach {\protect \cite{WiA93}},
 full curve: retarded approach of {\protect \cite{ScA01b}}.
 Offshell contributions to the nucleonic one-body current are
 included, too  {\protect \cite{ScA01c}}.
Experimental data from {\protect \cite{CrA96}} ($\bullet$), 
 {\protect \cite{ArG84}} (open box) and
 {\protect \cite{BlB95}} ($\circ$).}
\label{fig_gd}
\end{figure}

 As next, we discuss very briefly deuteron photodisintegration.
  The starting point of our consideration is the
   static approach of Wilhelm {\it et al.} \cite{WiA93} 
 which is based on the Bonn-OBEPR potential \cite{MaH87}.
 Similar to our  present approach, there is 
 no free parameter in the calculation of the 
  photodisintegration process in \cite{WiA93}. As is evident
 form Fig.\ \ref{fig_gd}, Wilhelm {\it et al.} 
 clearly fail in describing the  data. One obtains
  a considerable underestimation of the cross section in the $\Delta$ peak.
 Moreover, a dip structure around $90^{\circ}$ occurs at higher energies which
is not present in the data. 
 On the other hand,  these problems
 in the differential cross section  
 vanish almost completely in a retarded
 approach \cite{ScA01b,ScA01c}.  However, some discrepancies in  polarization
 observables like the linear photon asymmetry $\Sigma$ or the polarization
 $P_y(p)$ of the outgoing proton
  are still present, and  which need further consideration
 \cite{ScA01b,ScA01c}.

\section{Outlook}

A  very interesting topic to be studied in the  future
 is the exploration of the spin asymmetry 
 of the total cross section on the nucleon which
  determines the GDH-sum rule \cite{Ger65,DrH66} and  
which is at present under investigation  experimentally  \cite{AhA01}.  
 Due to the lack of a free neutron target, a measurement 
 on the deuteron is of specific significance because it may serve as an 
 effective  neutron target. However, the extraction of the
 neutron contributions
 relies on the basic assumption that
 final state interactions  and MEC can be neglected and that proton and neutron
 contribute incoherently. First, still preliminary
 results show that these assumptions are quite  crude. 
 In the future, we plan  to apply the present model to other
  reactions, especially 
 electrodisintegration.  Conceptually, 
 we have to  improve our hadronic interaction model. Moreover, additional
 d.o.f. like the Roper, the $D_{13}$ and the $S_{11}$  resonance should 
 be taken into account if one wants to consider higher energies.

\newpage

\end{document}